\documentclass[12pt,english]{article}
\usepackage[T1]{fontenc}
\usepackage[latin9]{inputenc}
\usepackage{amsmath}
\usepackage{amsthm}
\usepackage{setspace}
\makeatletter
\theoremstyle{plain}
\newtheorem{assumption}{\protect\assumptionname}
\theoremstyle{remark}
\newtheorem{rem}{\protect\remarkname}

\usepackage[letterpaper, left=1in, top=1in, right=1in, bottom=1in,nohead,includefoot, verbose, ignoremp]{geometry}
\usepackage{charter} 
\usepackage{enumerate} 
\usepackage{xy}\xyoption{all} \xyoption{poly} \xyoption{knot}  
\usepackage{latexsym,amssymb,amsmath,amsfonts,graphicx,color,fancyvrb,amsthm,enumerate,natbib,movie15}
\usepackage[pdftex,pagebackref=true]{hyperref}
\usepackage[svgnames,dvipsnames,x11names]{xcolor}
\usepackage{graphicx}
\usepackage{graphicx}
\usepackage{yhmath}
\usepackage{mathdots}
\usepackage{MnSymbol}
\hypersetup{
colorlinks,%
linkcolor=RoyalBlue2,  
urlcolor=Sienna4,   
citecolor=RoyalBlue2  
}

\def\beginmat{ \left( \begin{array} }
\def\endmat{ \end{array} \right) }

\setcitestyle{citesep={,}}

\makeatother

\usepackage{babel}
\providecommand{\assumptionname}{Assumption}
\providecommand{\remarkname}{Remark}

\begin{document}
\title{A Note on the Finite Sample Bias in Time Series Cross-Validation}
\author{Amaze Lusompa \thanks{I thank Johnson Oliyide for excellent research assistance. The views
expressed are those of the author and do not necessarily reflect the
positions of the Federal Reserve Bank of Kansas City or the Federal
Reserve System.}\\
Federal Reserve Bank of Kansas City\\
}
\maketitle
\begin{abstract}
It is well known that model selection via cross validation can be
biased for time series models. However, many researchers have argued
that this bias does not apply when using cross-validation with vector
autoregressions (VARs) or with time series models whose errors follow
a martingale-like structure. I show that even under these circumstances,
performing cross-validation on time series data will still generate
bias in general.
\end{abstract}
\pagebreak{}

\section*{1\quad{}Introduction}

Cross-validation (CV) is a widely used tool for model/hyperparameter
selection, model comparison, model averaging/combination, or to evaluate
the predictive ability of models and is an important tool in statistics,
econometrics, machine learning, and many other fields. It is considered
by many to be the gold standard of model selection. Though CV is flexible
and robust, it was developed for independent data. 

It is well known that model selection via CV can be biased for time
series data \citep{Altman1990,Hart1991,Opsomer2001}. It was first
argued by \cite{Burman1992} that this bias does not apply if the
residuals of the models follow a martingale difference sequence (MDS)
or martingale-like structure, and it has since become conventional
wisdom that CV will work well if this is the case (see for example
\cite{Burman1994,Racine2000,Arlot2010,Gelman2014,Bergmeir2018,Coulombe2022,Petropoulos2022}).
This argument has more recently picked up steam due to \cite{Bergmeir2018}
using a reformulated proof specific for Vector Autoregressions (VARs).
In this note I will show that CV with time series data will, in general,
be biased even if the residuals are MDS.

\section*{2\quad{}How Cross-Validation Works}

Let 
\[
y_{i}=\mu_{i}+\varepsilon_{i}\:for\:i=1,2,\ldots,T,
\]
where $y_{i}$ is the dependent variable, $\mu_{i}$ is the conditional
mean, and $\varepsilon_{i}$ is the residual. In the special case
of linear regressions
\[
y_{i}=x_{i}\beta+\varepsilon_{i}\:for\:i=1,2,\ldots,T,
\]
where $x_{i}$ is a $1\times p$ vector of regressors, $\beta$ is
a $p\times1$ vector of regression coefficients. The Leave-one-out
(LOO) residual for a regression model is calculated by estimating
$\beta$ for all but one observation, and then predicting the residual
for the left out observation. More formally, for observation $i$,
\[
\widetilde{\beta}_{-i}=(\sum_{j\neq i}x_{j}'x_{j})^{-1}\sum_{j\neq i}x_{j}'y_{j},\qquad\widetilde{\mu}_{-i}=x_{i}\widetilde{\beta}_{-i},\qquad\widetilde{\varepsilon}_{i}=y_{i}-\widetilde{\mu}_{-i},
\]
where $\widetilde{\beta}_{-i}$, $\widetilde{\mu}_{-i}$, and $\widetilde{\varepsilon}_{i}$
are the LOO estimates of $\beta$, $\mu_{i}$, and $\varepsilon_{i}$,
respectively. LOO CV is generally calculated as the mean-squared error
(MSE) of the LOO residuals, that is, $T^{-1}\sum_{i=1}^{T}\widetilde{\varepsilon}_{i}^{2}$,
though other loss functions could be used. The \cite{Burman1992}
and \cite{Bergmeir2018} arguments are with respect to the MSE case,
so that is the case that will be used throughout.

Define $\mathcal{A}{}_{T}$ as the set of all models being compared,
and $\alpha$ is the model index. Generally, the goal of CV (or any
model selection criterion) is to find the model with the smallest
average squared error (ASE) or mean average squared error (MASE).
The ASE is a distance or loss function between the true conditional
mean and the estimated conditional mean for model $\alpha$ and is
a measure of how well the conditional mean for model $\alpha$ approximates
the true conditional mean. A model with a smaller ASE means that model
is closer to the truth in the squared error loss sense. The ASE is
defined as
\[
\widetilde{L}_{T}(\alpha)=\frac{1}{T}\sum_{i=1}^{T}(\mu_{i}-\widetilde{\mu}_{-i}(\alpha))^{2}\qquad and\qquad L_{T}(\alpha)=\frac{1}{T}\sum_{i=1}^{T}(\mu_{i}-\hat{\mu}_{i}(\alpha))^{2}
\]
for the LOO and full sample estimates, respectively. 
\[
\widetilde{R}_{T}(\alpha)=E(\widetilde{L}_{T}(\alpha))\qquad and\qquad R_{T}(\alpha)=E(L_{T}(\alpha))
\]
are the LOO and full sample estimates of the MASE, respectively. CV
is considered to be a good model selection criterion in small samples
if CV can accurately choose the model with the smallest ASE or MASE
on average. 

\section*{3\quad{}Bias in Time Series Cross-Validation}

To illustrate the intuition behind CV, I will make the following assumption:
\begin{assumption}
The following conditions are satisfied for all models, $\alpha$,
in the set $A_{T}:$

(a) For all $i$, $E(\varepsilon_{i})=0$, $E(\varepsilon_{i}^{2})=\sigma^{2}$,
$E(\mu_{i}\varepsilon_{i})=0$.

(b) For each model, $\alpha$, and sample size T, $\frac{1}{T}\sum_{i=1}^{T}E(\mu_{i}-\hat{\mu}_{i}(\alpha))^{2}\approx\frac{1}{T}\sum_{i=1}^{T}E(\mu_{i}-\widetilde{\mu}_{-i}(\alpha))^{2}$.
\end{assumption}
\begin{rem}
This assumption should be uncontroversial and is standard \citep{Opsomer2001}.
Assumption 1(a) comprises of standard moment conditions for regressions.
Assumption 1(b), assuming the full sample MASE and the LOO MASE being
approximately the same, should hold in general unless some observations
have extremely high leverage.
\end{rem}
Note that since the following equation holds for the LOO residual
for model $\alpha$: $\widetilde{\varepsilon}_{-i}(\alpha)=y_{i}-\widetilde{\mu}_{-i}(\alpha)=\mu_{i}+\varepsilon_{i}-\widetilde{\mu}_{-i}(\alpha)$,
it can be shown via algebra that
\[
\widetilde{\varepsilon}_{-i}^{2}(\alpha)=\varepsilon_{i}^{2}+2(\mu_{i}-\widetilde{\mu}_{-i}(\alpha))\varepsilon_{i}+(\mu_{i}-\widetilde{\mu}_{-i}(\alpha))^{2}.
\]
Averaging over all observations, we have the following formula for
the CV MSE
\[
\frac{1}{T}\sum_{i=1}^{T}\widetilde{\varepsilon}_{-i}^{2}(\alpha)=\frac{1}{T}\sum_{i=1}^{T}\varepsilon_{i}^{2}+\frac{1}{T}\sum_{i=1}^{T}2\mu_{i}\varepsilon_{i}-\frac{1}{T}\sum_{i=1}^{T}2\widetilde{\mu}_{-i}(\alpha)\varepsilon_{i}+\frac{1}{T}\sum_{i=1}^{T}(\mu_{i}-\widetilde{\mu}_{-i}(\alpha))^{2}.
\]
CV is thought to perform well in finite samples if the expectation
of the CV MSE satisfies
\[
E(\frac{1}{T}\sum_{i=1}^{T}\widetilde{\varepsilon}_{-i}^{2}(\alpha))\approx\sigma^{2}+\frac{1}{T}\sum_{i=1}^{T}E(\mu_{i}-\hat{\mu}_{i}(\alpha))^{2}
\]
\citep{Altman1990,Hart1991,Opsomer2001}. If this formula holds, CV
will, on average, choose the model with the smallest MASE since the
only thing differentiating the expectation of the CV MSE across models
is the MASE for the respective models. By Assumption 1(a), $E(\frac{1}{T}\sum_{i=1}^{T}\varepsilon_{i}^{2})=\sigma^{2}$
and $E(\frac{1}{T}\sum_{i=1}^{T}2\mu_{i}\varepsilon_{i})=0$, so in
order for CV to perform well on average in finite samples, it essentially
needs to be the case that $E(\frac{1}{T}\sum_{i=1}^{T}\widetilde{\mu}_{-i}(\alpha)\varepsilon_{i})=0$.\footnote{CV could also perform well if $E(\frac{1}{T}\sum_{i=1}^{T}\widetilde{\mu}_{-i}(\alpha)\varepsilon_{i})\approx0$,
but the performance of CV would still depend on the magnitude of the
MASE, since $E(\frac{1}{T}\sum_{i=1}^{T}\widetilde{\mu}_{-i}(\alpha)\varepsilon_{i})\approx0$
could still affect the ordering of the models.} If $E(\frac{1}{T}\sum_{i=1}^{T}\widetilde{\mu}_{-i}(\alpha)\varepsilon_{i})\neq0$,
then the bias can change the ordering of the CV MSE and cause CV not
to choose the model with the smallest MASE on average. Note that outside
of razor's edge cases, in order for $E(\frac{1}{T}\sum_{i=1}^{T}\widetilde{\mu}_{-i}(\alpha)\varepsilon_{i})=0$,
it needs to be the case that $E(\widetilde{\mu}_{-i}(\alpha)\varepsilon_{i})=0$
for all $i$.

If the data is i.i.d. and $E(\varepsilon_{i}\big|x_{i}(\alpha))=0$
for all $\alpha$, then CV would be unbiased since 
\[
E[\widetilde{\mu}_{-i}(\alpha)\varepsilon_{i}]=E[x_{i}(\alpha)\widetilde{\beta}_{-i}(\alpha)\varepsilon_{i}]=E[x_{i}(\alpha)\varepsilon_{i}]E[\widetilde{\beta}_{-i}(\alpha)]=0,
\]
where the second equality is due to independence. For time series
data, it will generally be the case that $E(\widetilde{\mu}_{-i}(\alpha)\varepsilon_{i})\neq0$
since $\widetilde{\mu}_{-i}(\alpha)$ and $\varepsilon_{i}$ will
generally be correlated in finite samples, and contrary to conventional
wisdom in the literature, $\varepsilon_{i}$ being a MDS is not sufficient
to guarantee that $E(\widetilde{\mu}_{-i}(\alpha)\varepsilon_{i})=0$
for all $i$. 

As a counterexample, let us assume the true model is a stationary
AR(1) where 
\[
y_{i}=\rho y_{i-1}+\varepsilon_{i},
\]
$|\rho|<1$, $\rho\neq0$. Assume Assumption 1 still holds, so $\varepsilon_{i}$
is a mean zero random variable with finite variance. Also assume that
$\varepsilon_{i}$ is a MDS, that is, $E(\varepsilon_{i}|\varepsilon_{i-1},\varepsilon_{i-2},...)=0$
for all $i$. Lastly, let us also assume that $E[\widetilde{\mu}_{-i}(\alpha)\varepsilon_{i}]$
exists and is finite for all $i$ and $T$.\footnote{One could make more explicit distribution assumptions for $\varepsilon_{i}$
instead, but the argument would lose its generality.} For simplicity, let us also assume that the candidate model being
analyzed is the true model so that $\varepsilon_{i}(\alpha)=\varepsilon_{i}$
and $x_{i}(\alpha)=x_{i}$ and therefore 
\[
\widetilde{\mu}_{-i}(\alpha)\varepsilon_{i}=x_{i}(\alpha)\widetilde{\beta}_{-i}(\alpha)\varepsilon_{i}=x_{i}(\sum_{j\neq i}x_{j}'x_{j})^{-1}(\sum_{j\neq i}x_{j}'y_{j})\varepsilon_{i}.
\]
Since 
\[
y_{i}=\rho y_{i-1}+\varepsilon_{i}=\sum_{m=0}^{\infty}\rho^{m}\varepsilon_{i-m},
\]
then 
\[
E[\widetilde{\mu}_{-i}(\alpha)\varepsilon_{i}]=E[x_{i}(\alpha)\widetilde{\beta}_{-i}(\alpha)\varepsilon_{i}]=E[y_{i-1}(\sum_{j\neq i}y_{j-1}'y_{j-1})(\sum_{j\neq i}y_{j-1}'y_{j})\varepsilon_{i}]
\]
\[
=E\{[\sum_{m=0}^{\infty}\rho^{m}\varepsilon_{i-1-m}](\sum_{j\neq i}[\sum_{m=0}^{\infty}\rho^{m}\varepsilon_{j-1-m}]'[\sum_{m=0}^{\infty}\rho^{m}\varepsilon_{j-1-m}])^{-1}(\sum_{j\neq i}[\sum_{m=0}^{\infty}\rho^{m}\varepsilon_{j-1-m}]'[\sum_{m=0}^{\infty}\rho^{m}\varepsilon_{j-m}])\varepsilon_{i}\},
\]
which clearly will not be equal to zero in general for all $i$ since
the MDS assumption is not sufficient to guarantee it.\footnote{An exception would be when $i=T$. When $i=T$, $E[\widetilde{\mu}_{-i}(\alpha)\varepsilon_{i}]=0$
due to the MDS assumption that $E(\varepsilon_{T}|\varepsilon_{T-1},\varepsilon_{T-2},\ldots)=0$
and since $x_{T}=y_{T-1}$ and $\widetilde{\beta}_{-T}$ are functions
of $\varepsilon$ up to time $T-1$, $E[\widetilde{\mu}_{-i}(\alpha)\varepsilon_{i}]=E(E[\varepsilon_{i}|\widetilde{\mu}_{-i}(\alpha)])=0$.
More generally, the value of $E[\widetilde{\mu}_{-i}(\alpha)\varepsilon_{i}]$
can depend on $i$.}\textsuperscript{,}\footnote{Note that the bias goes away asymptotically under general conditions
(see for example Theorem 1 in \cite{Lusompa2025} for the proof).
Intuitively, this is because $\widetilde{\beta}_{-i}(\alpha)$ will
converge in probability to $\beta(\alpha)$, which is simply the convergence
of pseudo true parameters (see for example \cite{White1982,White1994}
and references therein for more details). Since $\beta(\alpha)$ is
a constant vector and assuming the standard moment condition $E[x_{i}(\alpha)\varepsilon_{i}]=0$
holds, $E[x_{i}(\alpha)\varepsilon_{i}\beta(\alpha)]=E[x_{i}(\alpha)\varepsilon_{i}]\beta(\alpha)=0$.} This is because even though the MDS assumption ensures the error
terms are uncorrelated, it allows for dependence in the error terms,
so the expectation of functionals of the error terms will not be zero
in general. In this specific counterexample, the bias is somewhat
reminiscent on the finite sample bias for autocovariances or VAR parameters
(see for example \cite{Kendall1954,Marriott1954,Shaman1988}) except
things are further complicated by the additional multiplication by
$x_{i}\varepsilon_{i}$.\footnote{For more general cases, the bias would be somewhat reminiscent of
the more general time series bias formula (see for example \cite{Anatolyev2005,Bao2007}).} The above counterexample assumes the true model is an AR(1) model,
and the candidate model being examined is the true model, but in general,
the bias will depend on the candidate model and the true data generating
process, so it will generally be the case that the bias can vary from
one candidate model to the next and can affect the model chosen by
CV. 

Since most time series models include lagged dependent variables and
are assumed to have a Wold type representation or a Wold type component
(see for example \cite{Priestley1981,Brockwell1991}), it will generally
be the case that $E(\widetilde{\mu}_{-i}(\alpha)\varepsilon_{i})\neq0$
in finite samples. So just like lagged dynamics in the dependent variable
cause bias in regression coefficients for time series models, it would
also cause bias in CV. If strict exogeneity was satisfied for the
true and candidate regressors, that is, for each candidate model $E(\varepsilon_{i}|x_{1}(\alpha),x_{1},\ldots,x_{T}(\alpha),x_{T})=0$
for all $i$ in addition to a martingale type assumption such as the
residuals satisfying $E(\varepsilon_{i}|\varepsilon_{j})=0$ for $i>j$,
then the bias would be zero and time series CV could carry on as if
the data were independent. Unfortunately, strict exogeneity is an
unrealistic assumption for time series that only applies in a small
number of situations at best \citep{Stock2007}. 

To clarify what \cite{Burman1992} and \cite{Bergmeir2018} do, their
arguments are with respect to what is known in the nonparametric literature
as the correlation effect in the variance of the MASE decomposition
\citep{Altman1990,Hart1991,Opsomer2001}. This is separate from the
bias that is coming from $\frac{1}{T}\sum_{i=1}^{T}\widetilde{\mu}_{-i}(\alpha)\varepsilon_{i}$.
\cite{Burman1992} and \cite{Bergmeir2018} assume that 
\[
E(\frac{1}{T}\sum_{i=1}^{T}\widetilde{\varepsilon}_{-i}^{2}(\alpha))=\sigma^{2}+\frac{1}{T}\sum_{i=1}^{T}E(\mu_{i}-\widetilde{\mu}_{-i}(\alpha))^{2},
\]
which we know is not true in general since $E[\frac{1}{T}\sum_{i=1}^{T}\widetilde{\mu}_{-i}(\alpha)\varepsilon_{i}]\neq0$
in general, even if the residuals follow a martingale-like structure.\footnote{\cite{Burman1992} write the proofs in terms of the full sample estimates
$E(\frac{1}{T}\sum_{i=1}^{T}\hat{\varepsilon}_{i}^{2}(\alpha))=\sigma^{2}+\frac{1}{T}\sum_{i=1}^{T}E(\mu_{i}-\hat{\mu}_{i}(\alpha))^{2}$,
as opposed to using the LOO estimates as I do, but the same arguments
still hold since $\frac{1}{T}\sum_{i=1}^{T}\hat{\varepsilon}_{i}^{2}(\alpha)$
can be decomposed the same way as $\frac{1}{T}\sum_{i=1}^{T}\widetilde{\varepsilon}_{-i}^{2}(\alpha)$
with $\widetilde{\mu}_{-i}(\alpha)$ being replaced by $\hat{\mu}_{i}(\alpha)$.
It is still the case that $E[\frac{1}{T}\sum_{i=1}^{T}\hat{\mu}_{i}(\alpha)\varepsilon_{i}]\neq0$
in general.}\textsuperscript{,}\footnote{\cite{Bergmeir2018} assume this formula is approximately true instead
of being exactly true.} \cite{Burman1992} and \cite{Bergmeir2018} then decompose $E(\mu_{i}-\widetilde{\mu}_{-i}(\alpha))^{2}$
via the standard bias-variance decomposition of the estimator. That
is, 
\[
E(\mu_{i}-\widetilde{\mu}_{-i}(\alpha))=\underbrace{E([\widetilde{\mu}_{-i}(\alpha)-E(\widetilde{\mu}_{-i}(\alpha)]^{2})}_{variance}+\underbrace{(E(\widetilde{\mu}_{-i}(\alpha))-\mu_{i})^{2}}_{squared\,bias}.
\]
What \cite{Burman1992} and \cite{Bergmeir2018} analyze is the impact
that errors have on the variance component in the bias-variance decomposition
of $E(\mu_{i}-\widetilde{\mu}_{-i}(\alpha))$ when they follow a martingale-like
structure versus when they do not. It is debatable whether one should
even care about this impact if what one actually cares about is choosing
the model with the smallest ASE or MASE since the magnitude of the
ASE and MASE already account for the impact or lack there from the
errors. It is also besides the point since even if there is no correlation
effect on the variance component in the bias-variance decomposition,
$E(\frac{1}{T}\sum_{i=1}^{T}\widetilde{\mu}_{-i}(\alpha)\varepsilon_{i})\neq0$
in general for finite samples, so CV with time series data will be
biased in general.

\section*{4\quad{}Conclusion}

I show that contrary to conventional wisdom, model errors following
a martingale-like structure is not sufficient for model selection
via CV to be unbiased for time series data. If one is concerned with
the finite sample bias, the block CV methods of \cite{Chu1991} and
\cite{Burman1994} which were designed to alleviate the impact of
the bias or alternative methods (see \cite{Opsomer2001} or \cite{Arlot2010}
for reviews) may be better alternatives in terms of finite sample
performance.

\bibliographystyle{chicago}
\bibliography{References}

@article{Anatolyev2005,
	author = {Stanislav Anatolyev},
	journal = {Econometrica},
	number = {3},
	pages = {983-1002},
	title = {GMM, GEL, Serial Correlation, and Asymptotic Bias},
	volume = {73},
	year = {2005}}

@article{Bao2007,
	author = {Yong Bao and Aman Ullah},
	journal = {Journal of Econometrics},
	number = {2},
	pages = {650-669},
	title = {The second-order bias and mean squared error of estimators in time-series models},
	volume = {140},
	year = {2007}}

@article{Gelman2014,
	author = {Andrew Gelman and Jessica Hwang and Aki Vehtari},
	journal = {Statistics and Computing},
	pages = {997--1016},
	title = {Understanding predictive information criteria for Bayesian models},
	volume = {24},
	year = {2014}}

@article{Petropoulos2022,
	author = {Fotios Petropoulos and Daniele Apiletti and Vassilios Assimakopoulos and Mohamed Zied Babai and Devon K. Barrow and Souhaib Ben Taieb and Christoph Bergmeir and Ricardo J. Bessa and Jakub Bijak and John E. Boylan and Jethro Browell and Claudio Carnevale and Jennifer L. Castle and Pasquale Cirillo and Michael P. Clements and Clara Cordeiro and Fernando Luiz Cyrino Oliveira and Shari De Baets and Alexander Dokumentov and Joanne Ellison and Piotr Fiszeder and Philip Hans Franses and David T. Frazier and Michael Gilliland and M. Sinan G{\"o}n{\"u}l and Paul Goodwin and Luigi Grossi and Yael Grushka-Cockayne and Mariangela Guidolin and Massimo Guidolin and Ulrich Gunter and Xiaojia Guo and Renato Guseo and Nigel Harvey and David F. Hendry and Ross Hollyman and Tim Januschowski and Jooyoung Jeon and Victor Richmond R. Jose and Yanfei Kang and Anne B. Koehler and Stephan Kolassa and Nikolaos Kourentzes and Sonia Leva and Feng Li and Konstantia Litsiou and Spyros Makridakis and Gael M. Martin and Andrew B. Martinez and Sheik Meeran and Theodore Modis and Konstantinos Nikolopoulos and Dilek {\"O}nkal and Alessia Paccagnini and Anastasios Panagiotelis and Ioannis Panapakidis and Jose M. Pav{\'\i}a and Manuela Pedio and Diego J. Pedregal and Pierre Pinson and Patr{\'\i}cia Ramos and David E. Rapach and J. James Reade and Bahman Rostami-Tabar and Micha{\l} Rubaszek and Georgios Sermpinis and Han Lin Shang and Evangelos Spiliotis and Aris A. Syntetos and Priyanga Dilini Talagala and Thiyanga S. Talagala and Len Tashman and Dimitrios Thomakos and Thordis Thorarinsdottir and Ezio Todini and Juan Ram{\'o}n Trapero Arenas and Xiaoqian Wang and Robert L. Winkler and Alisa Yusupova and Florian Ziel},
	journal = {International Journal of Forecasting},
	number = {3},
	pages = {705-871},
	title = {Forecasting: theory and practice},
	volume = {38},
	year = {2022}}

@article{Chu1991,
	author = {C.-K. Chu and J. S. Marron},
	journal = {The Annals of Statistics},
	number = {4},
	pages = {1906-1918},
	title = {Comparison of Two Bandwidth Selectors with Dependent Errors},
	volume = {19},
	year = {1991}}

@article{Kendall1954,
	author = {M. G. Kendall},
	journal = {Biometrika},
	number = {3/4},
	pages = {403--404},
	title = {NOTE ON BIAS IN THE ESTIMATION OF AUTOCORRELATION},
	volume = {41},
	year = {1954}}

@article{Marriott1954,
	author = {F. H. C. Marriott and J. A. Pope},
	journal = {Biometrika},
	number = {3/4},
	pages = {390-402},
	title = {Bias in the Estimation of Autocorrelations},
	volume = {41},
	year = {1954}}

@article{Shaman1988,
	author = {Paul Shaman and Robert A. Stine},
	journal = {Journal of the American Statistical Association},
	number = {403},
	pages = {842-848},
	title = {The Bias of Autoregressive Coefficient Estimators},
	volume = {83},
	year = {1988}}

@article{Altman1990,
	author = {N. S. Altman},
	journal = {Journal of the American Statistical Association},
	number = {411},
	pages = {749-759},
	title = {Kernel Smoothing of Data With Correlated Errors},
	volume = {85},
	year = {1990}}

@book{White1994,
	author = {Halbert White},
	publisher = {Cambridge University Press},
	title = {Estimation, inference and specification analysis},
	year = {1994}}

@article{Lusompa2025,
	author = {Amaze Lusompa},
	journal = {Federal Reserve Bank of Kansas City Working Paper No. RWP 25-12},
	title = {Regression Model Selection Under General Conditions},
	year = {2025}}

@article{Coulombe2022,
	author = {Philippe Goulet Coulombe and Maxime Leroux and Dalibor Stevanovic and St{\'e}phane Surprenant},
	journal = {Journal of Applied Econometrics},
	number = {5},
	pages = {920-964},
	title = {How is machine learning useful for macroeconomic forecasting?},
	volume = {37},
	year = {2022}}

@article{Arlot2010,
	author = {Sylvain Arlot and Alain Celisse},
	journal = {Statistical Survey},
	pages = {40-79},
	title = {A survey of cross-validation procedures for model selection},
	volume = {4},
	year = {2010}}

@article{Bergmeir2018,
	author = {Christoph Bergmeir and Rob J. Hyndman and Bonsoo Koo},
	journal = {Computational Statistics \& Data Analysis},
	pages = {70-83},
	title = {A note on the validity of cross-validation for evaluating autoregressive time series prediction},
	volume = {120},
	year = {2018}}

@article{Burman1992,
	author = {P. Burman and D. Nolan},
	journal = {Journal of Time Series Analysis},
	number = {3},
	pages = {189-207},
	title = {DATA-DEPENDENT ESTIMATION OF PREDICTION FUNCTIONS},
	volume = {13},
	year = {1992}}

@article{Opsomer2001,
	author = {Jean Opsomer and Yuedong Wang and Yuhong Yang},
	journal = {Statistical Science},
	number = {2},
	pages = {134-153},
	title = {Nonparametric Regression with Correlated Errors},
	volume = {16},
	year = {2001}}

@article{Burman1994,
	author = {Prabir Burman and Edmond Chow and Deborah Nolan},
	journal = {Biometrika},
	number = {2},
	pages = {351-358},
	title = {A Cross-Validatory Method for Dependent Data},
	volume = {81},
	year = {1994}}

@article{Racine2000,
	author = {Jeff Racine},
	journal = {Journal of Econometrics},
	number = {1},
	pages = {39-61},
	title = {Consistent cross-validatory model-selection for dependent data: hv-block cross-validation},
	volume = {99},
	year = {2000}}

@book{Priestley1981,
	author = {M. B. Priestley},
	publisher = {Academic Press},
	title = {Spectral Analysis and Time Series},
	year = {1981}}

@article{Hart1991,
	author = {Jeffrey D. Hart},
	journal = {Journal of the Royal Statistical Society. Series B},
	number = {1},
	pages = {173-187},
	title = {Kernel Regression Estimation With Time Series Errors},
	volume = {52},
	year = {1991}}

@article{White1982,
	author = {Halbert White},
	journal = {Econometrica},
	number = {1},
	pages = {1-25},
	title = {Maximum Likelihood Estimation of Misspecified Models},
	volume = {50},
	year = {1982}}

@book{Brockwell1991,
	author = {Peter J. Brockwell and Richard A. Davis},
	edition = {Second},
	publisher = {Springer-Verlag New York},
	title = {Time Series: Theory and Methods},
	year = {1991}}

@book{Stock2007,
	author = {James Stock and Mark Watson},
	editor = {3rd},
	publisher = {Addison Wesley Longman},
	title = {Introduction to Econometrics},
	year = {2007}}

\end{document}